\documentclass[
aps,
prd,
showpacs,
preprint,
tightenlines,
superscriptaddress,
nofootinbib,
floatfix,
a4paper]
{revtex4}
\usepackage{graphicx}
\begin{document}
\title{    Reexamining nonstandard interaction effects
\\         on supernova neutrino flavor oscillations
}
\author{        G.L.~Fogli}
\affiliation{   Dipartimento di Fisica
                and Sezione INFN di Bari\\
                Via Amendola 173, 70126 Bari, Italy\\}
\author{        E.~Lisi}
\affiliation{   Dipartimento di Fisica
                and Sezione INFN di Bari\\
                Via Amendola 173, 70126 Bari, Italy\\}
\author{        A.~Mirizzi}
\affiliation{   Dipartimento di Fisica
                and Sezione INFN di Bari\\
                Via Amendola 173, 70126 Bari, Italy\\}
\author{        D.~Montanino}
\affiliation{   Dipartimento di Scienza dei Materiali
                and Sezione INFN di Lecce\\
                Via Arnesano, 73100 Lecce, Italy\\}
\begin{abstract}
Several extensions of the standard electroweak model allow new
four-fermion interactions $\overline\nu_\alpha \nu_\beta \overline
f f$ with strength $\epsilon_{\alpha \beta}\, G_F$, where
($\alpha,\, \beta)$ are flavor indices. We revisit their effects
on flavor oscillations of massive (anti)neutrinos in supernovae,
in order to achieve,  in the region above the protoneutron star,
an analytical treatment valid for generic values of the neutrino
mixing angles $(\omega,\, \phi,\, \psi) = (\theta_{12},\,
\theta_{13},\, \theta_{23})$. Assuming that $\epsilon_{\alpha
\beta}\ll 1$, we find that the leading effects on the flavor
transitions occurring at high ($H$) and low ($L$) density along
the supernova matter profile can be simply embedded through the
replacements $\phi\to\phi+\epsilon_H$ and $\omega \to \omega +
\epsilon_L$, respectively, where $\epsilon_H$ and $\epsilon_L$ are
specific linear combinations of the $\epsilon_{\alpha \beta}$'s.
Similar replacements hold for eventual oscillations in the Earth
matter. From a phenomenological point of view, the most relevant
consequence is a possible uncontrolled bias ($\phi\to\phi +
\epsilon_H$) in the value of the mixing angle $\phi$ inferred by
inversion of supernova neutrino data. Such a drawback, however,
does not preclude the discrimination of the neutrino mass spectrum
hierarchy (direct or inverse) through supernova neutrino
oscillations.
\end{abstract}
\medskip
\pacs{
14.60.Pq, 
14.60.St, 
13.15.+g, 
97.60.Bw} 
\maketitle

\section{Introduction}

Atmospheric and solar neutrino data provide convincing evidence
for lepton flavor nonconservation  in the channels
$\nu_\mu\to\nu_\tau$ and $\nu_e\to\nu_{\mu,\tau}$, respectively
\cite{Revi}. The Maki-Nakagawa-Sakata-Pontecorvo hypothesis of
$\nu$ masses and mixings \cite{Maki} suffices to explain such
phenomena in terms of flavor oscillations \cite{Revi}, with no
need for additional neutrino properties or interactions.

However, new neutrino interactions are often predicted (together
with neutrino masses and mixings) by several extensions of the
standard electroweak theory, in the form of effective (low-energy)
four-fermion operators $\mathcal{O}_{\alpha\beta}\sim
\overline\nu_\alpha\nu_\beta \overline f f$  with strength
$G_{\alpha\beta}^f$, inducing either flavor-changing
($\alpha\neq\beta$) or flavor non-universal $(\alpha=\beta)$
neutrino transitions. Models with left-right symmetry \cite{LeRa}
or with supersymmetry with broken $R$-parity \cite{Rpar,Vi95}
provide widely studied realizations of such operators. Their net
effect in ordinary matter (with fermion density $N_f(x)$ at
position $x$) is to provide extra $\nu$ interaction energies
\cite{Wo78} proportional to the following dimensionless couplings
\begin{eqnarray}
\label{breakeps}
\epsilon_{\alpha\beta}(x)
  &\equiv&
  \sum_{f=e,u,d}G_{\alpha\beta}^f N_f(x)\Big/G_F\,N_e(x) \\
  &=& \epsilon_{\alpha\beta}^e  +
  \epsilon_{\alpha\beta}^u N_u(x)/N_e(x) +
  \epsilon_{\alpha\beta}^d N_d(x)/N_e(x) \\
  &=& \label{Ye} \epsilon_{\alpha\beta}^e+\epsilon_{\alpha\beta}^u
  - \epsilon_{\alpha\beta}^d+(\epsilon_{\alpha\beta}^u+
  2\epsilon_{\alpha\beta}^d)/Y_e(x)\ ,
\end{eqnarray}
where $\epsilon^f_{\alpha\beta}=G_{\alpha\beta}^f/G_F$, and
$Y_e(x)$ is the electron/nucleon number fraction.

As a consequence, in the flavor basis
$\nu_\alpha=(\nu_{e},\nu_\mu,\nu_\tau)$, the standard  interaction
hamiltonian in matter \cite{Wo78,Mi78}%
\footnote{Up to terms proportional to the unit matrix, and
discarding negligible one-loop corrections \protect\cite{Bo87}.}
\begin{equation}\label{Vabstd}
  H^\mathrm{std}_\mathrm{int}=\mathrm{diag}(V,\,0,\,0)\ ,\\
 \end{equation}
can be generalized as
\begin{equation}\label{Vabnonstd}
  (H_\mathrm{int})_{\alpha\beta}=(
  H_\mathrm{int}^\mathrm{std})_{\alpha\beta}+
  V\epsilon_{\alpha\beta}\ ,
\end{equation}
where
\begin{equation}
 \label{Vstd}
  V(x)=\sqrt{2}\,G_F\,N_e(x)\ .
\end{equation}
The interaction hamiltonian for antineutrinos is obtained by
replacing $G_{\alpha\beta}^f$ with $-(G^f_{\alpha\beta})^*$,
namely, $V$ with $-V$ and $\epsilon_{\alpha\beta}$ with
$\epsilon_{\alpha\beta}^*$.%
\footnote{Notice also that $\epsilon_{\alpha\beta}^* =
\epsilon_{\beta\alpha}$ for hermiticity of $H_\mathrm{int}$.}

Nonstandard neutrino interactions in matter, and their interplay
with the oscillation phenomenon, have been investigated in many
different contexts. An incomplete list includes analyses related
to the solar neutrino problem
\cite{Ro91,Gu91,Gu92,Ba91,Ri93,Fo94,Kr97,Ma98,Be98,BeKa,Be00,Gago,GuNu},
to the atmospheric neutrino anomaly \cite{Fo01}, to the LSND
oscillation results \cite{Be99}, to neutrinoless double beta decay
\cite{Doub}, to supernova neutrinos \cite{BeKa,Ku98}, to the
production or detection of laboratory neutrinos
\cite{Gr95,Jo98,Ot02}, and to future (very) long baseline projects
\cite{Ga01,Da01,Go01,Sa02,Hu01,Sc01,Hu02}. In a nutshell, these
works demonstrate that nonstandard neutrino interactions in matter
can (sometimes profoundly) modify the interpretation of current
and future oscillation searches, and that their possible effects
deserve particular attention, whenever inferences about standard
mass-mixing parameters are made from experimental data.

In this work, we revisit the effects of subleading nonstandard
four-fermion interactions on the flavor oscillation of massive and
mixed (anti)neutrinos in supernovae. We shall limit ourselves to
the region above the protoneutron star, where neutrinos propagate
freely, and all relevant fermion densities are
monotonically decreasing (roughly as the third power of the radius).%
\footnote{Possible flavor transitions induced by nonstandard
interactions near the neutrinosphere \cite{Va87,Vall} will be
commented at the end of Sec.~II.}
For this case (previously studied, e.g., in \cite{BeKa,Ku98}) we
aim to a completely analytical treatment, valid for generic values
of the neutrino mixing angles and for both the direct and inverse
mass spectrum hierarchies allowed by the current $\nu$
phenomenology. Such treatment has been recently proposed in
\cite{Sour} for the case of {\em standard\/} supernova $\nu$
oscillations.

In the present work, we  make a few simplifying assumptions about
the nonstandard couplings $\epsilon_{\alpha\beta}$: (1) they are
taken as real (either positive or negative), so that
$\epsilon_{\alpha\beta}=\epsilon_{\beta\alpha}$; (2) their
possible variation with $x$ through $Y_e$ [see Eq.~(\ref{Ye})] is
neglected;%
\footnote{Detailed supernova simulations show that $Y_e\sim 0.5$
above the protoneutron star. Significant variations (reductions)
of $Y_e$ are confined to regions near the neutrinosphere during
the cooling phase \cite{Ylow}.}
and (3) they are assumed to be small,
\begin{equation}\label{epsab}
  |\epsilon_{\alpha\beta}| \ll 1\ ,
\end{equation}
say, $|\epsilon_{\alpha\beta}| <\mathrm{few}\times 10^{-2}$, so
that terms of $O(\epsilon^2)$ can be safely neglected in the
calculations.

Under the above assumptions, we find that the main effect of
nonstandard interactions on supernova neutrino oscillations is to
induce a shift in the relevant mixing angles, as compared to the
standard case. Such a shift is discussed first in the two-neutrino
case (Sec.~II) and then in the three-neutrino case (Sec.~III).
Nonstandard effects in the Earth matter are briefly discussed in
Sec.~IV. Phenomenological consequences are illustrated  in Sec.~V.
Conclusions are drawn in Sec.~VI.

A final remark is in order. The assumption in Eq.~(\ref{epsab}) is
consistent with phenomenological upper bounds
\cite{BeKa,Be00,Be99,Gr00}, derived from charged lepton processes
under $SU(2)$ symmetry assumptions. Such bounds are typically of
$O(10^{-1})$ for the (differences of) diagonal couplings and of
$O(10^{-2})$ for the off-diagonal ones; only the upper limit on
$\epsilon_{e\mu}$ is much more severe [$O(10^{-5})$].  However, if
allowance is made for substantial $SU(2)$ breaking, and if the
analysis is performed in a more model-independent way, the
previous constraints \cite{BeKa,Be00,Be99,Gr00} can be
significantly relaxed \cite{Ro01,Ra01}. Accordingly, scenarios
with (some) large $\epsilon_{\alpha\beta}$'s [$\sim
\mathrm{few}\times 10^{-1}$] can be built (see, e.g.,
\cite{Ro01,Ru95}), but they will not be considered in this paper.

\section{Two-neutrino transitions}

In this section, we study the effects of nonstandard interactions
on $2\nu$ oscillations, in the region well above the
neutrinosphere. We show that the $2\nu$ evolution equation can be
formally cast in an effective (and solvable) standard form
($\epsilon_{\alpha\beta} = 0$), through a proper redefinition of
the neutrino mixing angle. The traditional formulae, derived
through resonance-condition arguments \cite{Ma98,Be98,BeKa,Ku98}
(inapplicable for mixing angles $\gtrsim\pi/4$, see \cite{Sour}
and references therein), are shown to be recovered in the limit of
small mixing. At the end of this section, we also comment on a
different mechanism that might induce nonstandard flavor
transitions near the neutrinosphere.

In the flavor basis, the two-family evolution equation for
free-streaming neutrinos can be written as
\begin{equation}\label{2nuevol}
  i\, \frac{d}{dx}\left(\begin{array}{c}\nu_e\\ \nu_a\end{array}\right) =
  [H_\mathrm{kin}(\theta)+H_\mathrm{int}(x)]
  \left(\begin{array}{c}\nu_e\\ \nu_a \end{array}\right)\ ,
\end{equation}
where $a=\mu$ or $\tau$, and $H = H_\mathrm{kin}(\theta)+
H_\mathrm{int}(x)$ is the total hamiltonian, split into kinetic
and interaction energy terms. The kinetic term can be expressed in
terms of the neutrino mixing angle $\theta$, of the squared mass
difference $\Delta m^2=m^2_2-m^2_1$, and of the energy $E$ as
\begin{equation}\label{2Hkin}
  H_\mathrm{kin}=U(\theta)
  \left(
  \begin{array}{cc}
    -k/2 & 0 \\
    0 & +k/2
  \end{array}
  \right)
  U^\dag(\theta)\ ,
\end{equation}
where $k=\Delta m^2/2E$ is the neutrino oscillation wave number,
and
\begin{equation}\label{Utheta}
  U(\theta)=\left(
  \begin{array}{cc}
  \cos\theta  & \sin\theta\\
  -\sin\theta & \cos\theta
  \end{array}
  \right)\
\end{equation}
is the mixing matrix. The interaction term can be written as
\begin{eqnarray}
  H_\mathrm{int}(x)
  &=&
  V(x)
  \left(
  \begin{array}{cc}
    1+\epsilon_{ee} & \epsilon_{ea}  \\
    \epsilon_{ea}   & \epsilon_{aa}
  \end{array}
  \right)\label{2Hint1}\\
  &\equiv&
  \frac{V(x)}{2}
  \left(
  \begin{array}{cc}
    1+\epsilon_{ee}-\epsilon_{aa} & 2\epsilon_{ea}  \\
    2\epsilon_{ea}   & -1+\epsilon_{ee}+\epsilon_{aa}
  \end{array}
  \right)\ ,
  \label{2Hint2}
\end{eqnarray}
where we have made $H_\mathrm{int}$ traceless in the last
equation, $V(x)$ being defined as in Eq.~(\ref{Vstd}). For
$\epsilon_{\alpha\beta}=0$ (standard case), the interaction term
reduces to
\begin{equation}
\label{Hstd2nu} H^\mathrm{std}_\mathrm{int} =
\mathrm{diag}(+1,\,-1) \cdot V(x)/2 \ .
\end{equation}

The term $H_\mathrm{int}$ in Eq.~(\ref{2Hint2}) is diagonalized
through
\begin{equation}\label{2Hintdiag}
  H_\mathrm{int}= U(\eta)^\dag\, \mathrm{diag}(+1,\,-1)\, U(\eta)
  \cdot f\,V(x)/2\ ,
\end{equation}
where
\begin{equation}\label{factor}
  f = \sqrt{(1+\epsilon_{ee}-\epsilon_{aa})^2+4\epsilon_{ea}^2}\ ,
\end{equation}
while $U(\eta)$ is defined as in Eq.~(\ref{Utheta}), but with
argument
\begin{eqnarray}
  \eta &=& \frac{1}{2}
  \arctan \frac{2\epsilon_{ea}}{1+\epsilon_{ee}-\epsilon_{aa}}
  \label{eta}\\
       &=& \epsilon_{ea} + O(\epsilon^2)
  \label{etaeps}\ ,
\end{eqnarray}
where terms of $O(\epsilon^2)$ can be neglected under the
assumption in Eq.~(\ref{epsab}).

Equation~(\ref{2Hintdiag}) differs from Eq.~(\ref{Hstd2nu}) in two
respects: (a) the potential $V$ is multiplied by a factor $f$, and
(b) there is an additional rotation matrix $U({\eta}) =
U(\epsilon_{ea}) + O(\epsilon^2)$. The point (a) is equivalent to
rescale the supernova density by less than a few percent
[$f=1+O(\epsilon)$], and does not lead to realistically measurable
consequences; in fact, the difference $f-1$ is conceivably smaller
than the uncertainties associated to the normalization and shape
of inferred or simulated supernova density profiles; moreover, the
oscillation physics depends on $V$ mainly through its logarithmic
derivative \cite{Sour}. The point (b) can instead be important
and, as we shall see, it leads to a bias in the ``effective mixing
angle'' governing flavor transitions along the supernova matter
profile.

Summarizing the above discussion, the leading nonstandard effect
on $2\nu$ propagation in supernovae is governed by the
off-diagonal coupling $\epsilon_{ea}$, and amounts to write
$H_\mathrm{int}$ in the form
\begin{equation}\label{2Hinteps}
  H_\mathrm{int}(x)\simeq
  \frac{V(x)}{2}\;
  U^\dag(\epsilon_{ea})
    \left(
    \begin{array}{cc}
    +1 & 0 \\
    0 & -1
    \end{array}
    \right)
  U(\epsilon_{ea})\ .
\end{equation}
It is then easy to check that, in the new basis
\begin{equation}\label{tildenu}
  \left(
  \begin{array}{c}
  \tilde{\nu}_e\\
  \tilde{\nu}_a
  \end{array}
  \right)=U(\epsilon_{ea})
  \left(
  \begin{array}{c}
  \nu_e\\
  \nu_a
  \end{array}
  \right)\ ,
\end{equation}
the neutrino evolution equation in matter [Eq.~(\ref{2nuevol})]
takes an effective ``standard'' form,
\begin{equation}\label{2nuevoleff}
  i\, \frac{d}{dx}\left(\begin{array}{c}\tilde\nu_e\\
  \tilde\nu_a\end{array}\right) =
  [H_\mathrm{kin}(\tilde\theta)+H_\mathrm{int}^\mathrm{std}(x)]
  \left(\begin{array}{c}\tilde\nu_e
  \\ \tilde\nu_a \end{array}\right)\ ,
\end{equation}
where $\tilde{\theta}=\theta+\epsilon_{ea}$. Roughly speaking,
nonstandard effects can be ``rotated away'' from $H_\mathrm{int}$
(which becomes $H_\mathrm{int}^\mathrm{std}$) and ``transferred''
to $H_\mathrm{kin}$ through the replacement
$\theta\to\tilde{\theta}=\theta+\epsilon_{ea}$. Since accurate
analytical solutions of Eq.~(\ref{2nuevoleff}) exist [expressed,
e.g., in terms of the survival probability
$P(\tilde{\nu}_e\to\tilde{\nu}_e)$], the task is reduced to a
proper back-rotation of known results (applied to
$\tilde{\nu}_\alpha$) to the physical flavor basis $\nu_\alpha$.

To do so, one defines the usual mixing angle in matter
$\tilde{\theta}_m$ diagonalizing the total hamiltonian in
Eq.~(\ref{2nuevoleff}),
\begin{equation}\label{thetatildem}
  \cos2\tilde{\theta}_m=\frac
  {\cos 2\tilde{\theta}-V/k}
  {\sqrt{(\cos 2\tilde{\theta}-V/k)^2+\sin^2 2\tilde{\theta}}}
\end{equation}
and the associated basis of effective mass eigenstates in matter
$\tilde{\nu}_{m,i}$,
\begin{equation}\label{nutildei}
  \tilde{\nu}_\alpha = U(\tilde{\theta}_m)\,\tilde{\nu}_{m,i}\ .
\end{equation}

The formal solution of Eq.~(\ref{2nuevoleff}), evolving the state
$(\nu_\alpha)$ from the starting point ($x=x_o$)  to the detection
point ($x=x_d$), can then be factorized as
\begin{eqnarray}\label{facttilde}
  \left(
  \begin{array}{c}
  \nu_e\\
  \nu_a
  \end{array}
  \right)_d &=&
  U(\theta)\cdot T_V \cdot T_S(\tilde{\theta})\cdot
  U^\dag(\tilde{\theta}_m^o)
  U(\epsilon_{ea})
  \left(
  \begin{array}{c}
  \nu_e\\
  \nu_a
  \end{array}
  \right)_o \\
  \label{fact}
  &=&
  U(\theta)\cdot T_V \cdot T_S(\tilde{\theta})\cdot
  U^\dag(\tilde{\theta}_m^o-\epsilon_{ea})
  \left(
  \begin{array}{c}
  \nu_e\\
  \nu_a
  \end{array}
  \right)_o\ ,
\end{eqnarray}
where, from right to left, $U(\epsilon_{ea})$ rotates the initial
state $\nu_\alpha$ to the $\tilde\nu_\alpha$ basis of
Eq.~(\ref{2nuevoleff}), $U^\dag(\tilde{\theta}_m^o)$ rotates
$\tilde\nu_\alpha$ to the initial basis of mass eigenstates
$\tilde{\nu}_{m,j}^o$ in matter, $T_S(\tilde{\theta})$ embeds the
transitions $\tilde{\nu}_{m,j}^o\to\tilde{\nu}_{m,i}$ up to the
supernova surface (where $V=0$ and $\tilde{\nu}_{m,i}=\nu_i$),
$T_V$ further propagates the $\nu_i$'s in vacuum until they arrive
at the detector (up to Earth matter effects), and a final rotation
$U(\theta)$ brings the $\nu_i$ states back to the physical flavor
basis $\nu_\alpha$. Equation~(\ref{fact}), obtained by grouping
the first two rotations, represents the formal solution of the
$\nu$ evolution equation, in a format analogous to the standard
case (i.e., initial rotation, propagation in matter, propagation
in vacuum, final rotation).

One can then average out unobservable oscillating terms (mainly
associated to $T_V$), so as to propagate classical (real)
probabilities $P_{\alpha\beta}$ rather than complex amplitudes.
The averaging can be accomplished \cite{KuPa} by replacing in
Eq.~(\ref{fact}) each entry in the matrices $U(\theta)$,
$T_S(\tilde{\theta})$, and
$U^\dag(\tilde{\theta}_m^o-\epsilon_{ea})$, with its squared
modulus. In our case, the initial (high density) condition
$V(x_o)/k\gg 1$  leads to $\sin^2(\tilde{\theta}_m^o -
\epsilon_{ea}) \simeq 1 + O(\epsilon_{ea}^2)$, effectively
cancelling any ``memory'' of the initial point $x_o$ above the
protoneutron star. Therefore, the electron neutrino survival
probability finally reads%
\footnote{ We remind the reader that $P_{ee}$ is the most relevant
quantity in supernova $\nu$ oscillations, due to the practical
indistinguishability of $\nu_\mu$ and $\nu_\tau$ in the initial
and final states \cite{KuPa}. See, however, Ref.~\cite{Akhm} for a
recent discussion of possible differences between $\nu_\mu$ and
$\nu_\tau$ fluxes at the origin.}
\begin{equation}\label{Pee2nufact}
    P_{ee} =
    \left(
    \begin{array}{cc}
    1 & 0
    \end{array}
    \right)
    \left(
    \begin{array}{cc}
    \cos^2\theta & \sin^2\theta \\
    \sin^2\theta & \cos^2\theta
    \end{array}
    \right)
    \left(
    \begin{array}{cc}
    1-P_c(\tilde{\theta}) & P_c(\tilde{\theta}) \\
    P_c(\tilde{\theta}) & 1-P_c(\tilde{\theta})
    \end{array}
    \right)
    \left(
    \begin{array}{cc}
    0 & 1 \\
    1 & 0
    \end{array}
    \right)
    \left(
    \begin{array}{c}
    1 \\
    0
    \end{array}
    \right)\ ,
\end{equation}
namely,
\begin{equation}\label{Pee2nu}
  P_{ee}=\cos^2\theta\,P_c(\tilde{\theta})+
  \sin^2\theta[1-P_c(\tilde{\theta})]\ ,
\end{equation}
where we have adopted the usual notation for the square moduli in
$T_S$, in terms of the so-called crossing probability $P_c$ (see,
e.g., \cite{KuPa}).

Equation~(\ref{Pee2nu}) is formally analogous to the one in the
standard case ($\epsilon_{ea}=0$), modulo the replacement
$\theta\to\tilde{\theta} = \theta + \epsilon_{ea}$ in $P_c$. We
recall that, in the standard case, $P_c(\theta)$ is accurately
described  by a simple analytical expression,
\begin{equation}\label{Pcstd}
  P^\mathrm{std}_c(\theta)=\frac
  {e^{2\pi r k\cos^2\theta}-1}
  {e^{2\pi r k}-1}\ ,
\end{equation}
where
\begin{equation}\label{rscale}
r = -\left( \frac{d \ln V}{dx} \right)_{x=x_p}^{-1}
\end{equation}
is the density scale factor, to be evaluated at the point $x_p$
where the potential equals the wave number \cite{Sour},
\begin{equation}\label{Veqk}
  V(x_p)=k\ .
\end{equation}
The above prescription for $P^\mathrm{std}_c(\theta)$, inspired by
the condition of maximum violation of adiabaticity
\cite{Alex,Petc,Sour} (more general than the traditional condition
of resonance \cite{KuPa}) and by the double-exponential
representation for the crossing probability \cite{KrPe}, holds
accurately for generic supernova density profiles, in the whole
$2\nu$ mass-mixing parameter space \cite{Sour}.

In the presence of nonstandard interactions, the previously
discussed replacement $\theta\to\theta+\epsilon_{ea}$ leads then
to the following analytical form for $P_c$:
\begin{equation}\label{Pceps}
  P_c(\tilde{\theta)}=\frac
  {e^{2\pi r k\cos^2(\theta+\epsilon_{ea})}-1}
  {e^{2\pi r k}-1}\ ,
\end{equation}
where $r$ takes the same value as in Eq.~(\ref{rscale}), the
position of $x_p$ [Eq.~(\ref{Veqk})] being unaffected by the shift
$\theta\to\theta+\epsilon_{ea}$.%
\footnote{The definition of $r(x_p)$ through Eqs.~(\ref{rscale})
and (\ref{Veqk}) is $\theta$-independent.}

Equations (\ref{Pee2nu}) and  (\ref{Pceps}) represent our main
result for the two-flavor neutrino case. They provide a compact
description of nonstandard interaction effects on $P_{ee}$, valid
for generic values of $\theta$ and $\Delta m^2$ (provided that
$\epsilon_{ea}\ll 1$). The calculation of $P_{ee}$ for
antineutrinos is strictly analogous: given the analytical
expression reported in Sec.~II.C of \cite{Sour}, one has simply to
replace $\theta$ with $\theta + \epsilon_{ea}$
within $P_c(\bar\nu)$.%
\footnote{Notice that, {\em outside} $P_c$, the angle $\theta$ is
unshifted, as in the coefficients $\sin^2\theta$ and
$\cos^2\theta$ of Eq.~(\ref{Pee2nu}).}

It can easily be shown that Eq.~(\ref{Pceps}) generalizes previous
results about nonstandard interaction effects, valid in a more
limited range of applicability. The traditional analytical
approach to $P_c$ localizes the nonadiabatic transitions at the
so-called ``resonance'' point $x_r$, where mixing in matter is
maximal ($\sin2\theta_m(x_r)=1$) \cite{Ma98,Be98,BeKa,Ku98}. The
resonance approach is certainly valid at small $\theta$, where it
should coincide with ours. Indeed, in the resonance approach, the
standard $(\epsilon_{ea}=0)$ Landau-Zener form for $P_c$ at small
mixing ($\ln P_c\propto \theta^2$) \cite{KuPa} is modified as $\ln
P_c \propto \theta^2 (1 + 2\epsilon_{ea}\cot 2\theta)^2 \simeq
(\theta+\epsilon_{ea})^2$ \cite{Ma98,Be98,BeKa,Ku98}, in agreement
with our prescription $\theta\to\theta+\epsilon_{ea}$. For
$\theta$ approaching $\pi/4$, however, the resonance condition
generally fails to localize the right point $x$ for the evaluation
of $r(x)$, and becomes eventually not applicable for
$\theta>\pi/4$ (both in the standard \cite{Sour} and in the
nonstandard case).
 Conversely, Eq.~(\ref{Pceps}) has
no such applicability restrictions. Finally, in the subcase
$\theta\equiv 0$, we also recover from Eq.~(\ref{Pceps}) the
double-exponential form for $P_c$ found analytically in
\cite{Gu92} in the scenario of {\em massless\/} (solar) neutrinos
with nonstandard interactions. Therefore, Eq.~(\ref{Pceps})
correctly generalizes previous limiting cases.

A final comment is in order. Additional nonstandard flavor
transitions (different from those described in this section) might
independently occur in inner layers \cite{Va87,Vall}, as a result
of a decrease of $Y_e(x)$ to $O(10^{-2})$, localized just above
the neutrinosphere during the cooling phase \cite{Ylow,Vall}.  The
$\epsilon_{\alpha\beta}(x)$'s can then locally increase to $O(1)$
[see Eq.~(\ref{breakeps})], making the diagonal and off-diagonal
terms in the interaction hamiltonian [Eq.~(\ref{2Hint1})]
comparable in magnitude, and dominant (due to the very high matter
density) with respect to the kinetic terms [Eq.~(\ref{2Hkin})]. As
a consequence, near the neutrinosphere (and thus prior to the
flavor oscillations discussed in this paper), nonstandard
interactions might induce additional flavor transitions of
(effectively massless) $\nu$ and $\bar\nu$ during the cooling
phase \cite{Va87,Vall}. Our results, applicable to the region of
decreasing fermion density above the protoneutron star (where
$Y_e\sim 0.5$) should be thus considered as complementary to the
possible independent effects \cite{Va87,Vall} possibly induced by
a local (preceding) drop of $Y_e$. A combination of the two
nonstandard interaction effects (still lacking in the literature,
to our knowledge) is, however, beyond
the scope of this paper.%
\footnote{We note in passing that, being confined to a region just
above the neutrinosphere (where collective and collision effects
can still be non negligible), the additional transitions at low
$Y_e$ might require a more advanced description than the
free-particle formalism used in \cite{Vall} (e.g., in terms of
kinetic and density matrix equations  \cite{Deco}).}

\section{Three-neutrino transitions}

In the $3\nu$ case, the interaction term in the Hamiltonian is
given by Eq.~(\ref{Vabnonstd}), while the kinetic term
$H_\mathrm{kin}$ is
\begin{equation}\label{3Hkin}
  H_\mathrm{kin} = U(\omega,\,\phi,\,\psi)
  \left(
  \begin{array}{ccc}
  -k_L/2  & 0 & 0 \\
   0 & +k_L/2 & 0 \\
   0 &  0 &    k_H
  \end{array}
  \right)
  U^\dag(\omega,\,\phi,\,\psi)\ ,
\end{equation}
where $k_L=\delta m^2/2E$ and $k_H=m^2/2E$ ($\delta m^2$ and $m^2$
being the ``solar'' and ``atmospheric'' neutrino squared mass
differences, see \cite{Revi,Sour}), while the $3 \times 3$ mixing
matrix $U(\omega,\,\phi,\,\psi)$ is factorized (discarding a
possible CP violating phase) into three real rotations
\cite{KuPa,Ku88}
\begin{equation}\label{U3nu}
U(\omega,\,\phi,\,\psi)=U_{23}(\psi)U_{13}(\phi)U_{12}(\omega)\ ,
\end{equation}
where $(\omega,\phi,\psi)=(\theta_{12},\theta_{13},\theta_{23})$.
The subscripts $H$ and $L$ remind that two transitions can occur
along the supernova profile when $V\sim O(k_H)$ and $V\sim
O(k_L)$, namely, at high ($H$) and low ($L$) density,
respectively.

The phenomenological assumption of hierarchical squared mass
differences,
\begin{equation}\label{hier}
  \delta m^2 \ll m^2 \longleftrightarrow k_L \ll k_H\ ,
\end{equation}
together with Eq.~(\ref{epsab}), makes the $3\nu$ dynamics
factorizable into two (almost decoupled) $2\nu$ subsystems for the
$H$ and $L$ transitions, in the same way as for the standard case
\cite{KuPa}.

To isolate the dynamics of the $H$ transition, one rotates the
starting neutrino (flavor) basis by $U_{23}^\dag(\psi)$, and
extracts the submatrix with indices $(1,\,3)$ \cite{KuPa,Ku98}.
The effective $2\nu$ sub-dynamics is then governed by
$(k,\theta)\equiv(k_H,\phi)$, and the analogous of $\epsilon_{ea}$
is given by the off-diagonal (nonstandard) term, which reads
\begin{equation}\label{epsH}
\epsilon_H=\epsilon_{e\mu}\,\sin\psi+\epsilon_{e\tau}\,\cos\psi\ .
\end{equation}
Using the results of the previous section and those of
Ref.~\cite{Sour}, the crossing probability for the $H$ transition
is thus described in terms of an effective $(1,3)$ mixing angle
$\tilde{\phi}=\phi+\epsilon_H$, so that
\begin{equation}\label{PHeps}
 P_H=\frac{e^{2\pi r_H k_H \cos^2(\phi+\epsilon_H)}-1}
 {e^{2\pi r_H k_H}-1}\ ,
\end{equation}
$r_H$ being the density scale factor [Eq.~(\ref{rscale})]
evaluated at the point $x_H$ where $V(x_H)=k_H$.

Analogously, to isolate the dynamics of the $L$ transition, one
rotates the starting neutrino (flavor) basis by
$U_{13}^\dag(\phi)U_{23}^\dag(\psi)$, and then extracts the
submatrix with indices $(1,\,2)$. The effective $2\nu$
sub-dynamics is then governed by $(k,\theta)\equiv(k_L,\omega)$
and by the off-diagonal (nonstandard) term, which now reads%
\footnote{An analogous result for $\epsilon_L$ has been recently
reported in the context of solar neutrinos \protect\cite{GuNu}.
The expression of $\epsilon_L$ for supernova neutrinos, in the
subcase $\epsilon_{\mu\tau} = \epsilon_{\mu\mu} =
\epsilon_{\tau\tau} = 0$, is reported in \protect\cite{Ku98}.}
\begin{eqnarray}\label{epsL}
\epsilon_L &=&
\cos\phi(\epsilon_{e\mu}\cos\psi-\epsilon_{e\tau}\sin\psi)
\nonumber\\&+&
\sin\phi[(\epsilon_{\tau\tau}-\epsilon_{\mu\mu})\sin\psi\cos\psi-
\epsilon_{\mu\tau}\cos2\psi]\ .
\end{eqnarray}
The crossing probability for the $L$ transition is thus described
in terms of an effective $(1,2)$ mixing angle
$\tilde{\omega}=\omega+\epsilon_L$, so that
\begin{equation}\label{PLeps}
 P_L=\frac{e^{2\pi r_L k_L \cos^2(\omega+\epsilon_L)}-1}
 {e^{2\pi r_L k_L}-1}\ ,
\end{equation}
$r_L$ being evaluated at the point $x_L$ where $V(x_L)=k_L$. The
same shift ($\omega\to\omega+\epsilon_L$) applies to the
expression of $P_L$ for antineutrinos (as given in \cite{Sour}).

The above discussion implicitly assumes direct hierarchy of
neutrino masses ($m_{1,2}<m_3$). The case of inverse hierarchy
($m_3<m_{1,2}$) is, however, strictly analogous; indeed, the
symmetry arguments used in \cite{Sour} to derive $P_{ee}$
analytically in the case of inverse hierarchy (in terms of
$P_{ee}$ for direct hierarchy) are unaffected by the presence of
nonstandard interactions. Therefore, the prescription remains the
same, irrespective of the spectrum type: take the standard
expressions of $P_{ee}$ from \cite{Sour}, and replace
$\omega\to\omega +\epsilon_L$ and $\phi\to\phi+\epsilon_H$ in the
crossing probabilities for both neutrinos and antineutrinos.
Embedding nonstandard interaction effects through appropriate
shifts of $\omega$ and $\phi$ within $P_L$ and $P_H$,
respectively, represents our main results for the case of
three-flavor mixing of supernova neutrinos (up to Earth matter
effects).

\section{Earth matter effects}

In the standard case $(\epsilon_{\alpha\beta}=0)$, it is well
known that possible Earth matter effects before supernova $\nu$
detection can be embedded through a specific transition
probability $P_E=P(\nu_2\to\nu_e)$ (see \cite{Sour,Di00,Lu01} and
references therein). In the hierarchical approximation
[Eq.~(\ref{hier})], $P_E$ is basically
a function of the $L$-transition parameters $(\omega,\delta m^2)$ only.%
\footnote{There is a subleading dependence on $\phi$ through the
renormalization $V\to V\cos^2\phi$, which can be safely neglected
for supernova neutrinos, given the smallness of $\phi$
\protect\cite{Revi} and unavoidable normalization uncertainties of
$V$.}

The dependence of $P_E$ on $\omega$ is both explicit (through an
$\omega$-rotation in vacuum), and implicit [through the mixing
angle in the Earth matter, $\omega_m=\omega_m(\omega)$], so that
\begin{equation}\label{PEstd}
 P_E=P_E(\omega,\, \omega_m(\omega),\, \delta m^2),\
 \mathrm{standard\
 case}\ .
\end{equation}
In particular, Sec.~IV of \cite{Sour} reports known
representations of the function $P_E(\omega,\, \omega_m(\omega),\,
\delta m^2)$ in the case of Earth mantle (+ core) crossing, in the
same notation as in the present paper.

In the nonstandard case, since it is $\epsilon_{\alpha\beta}\neq
0$ only in {\em matter}, the previously found shift
$\omega\to\omega + \epsilon_L$ applies only to $\omega_m$, namely,
\begin{equation}\label{PEnonstd}
 P_E=P_E(\omega,\, \omega_m(\omega+\epsilon_L),\, \delta m^2),\
 \mathrm{nonstandard\
 case}\ ,
\end{equation}
and similarly for antineutrinos.%
\footnote{In principle, the value of $\epsilon_L$ to be used in
Eq.~(\ref{PEnonstd}) might be slightly different from the one in
Eq.~(\ref{PLeps}), due to different (average) values of $Y_e$ in
the Earth and in the supernova [see Eq.~(\ref{Ye})].}
We have also checked the above prescription by explicitly
repeating the derivation of $P_E$ in the cases of one-layer or
two-layer model for the Earth crossing (omitted).

\section{Discussion}

Let us discuss in more detail the effects of nonzero $\epsilon_L$
and $\epsilon_H$ on supernova neutrino oscillations, by
considering the phenomenologically interesting case with
$m^2=3\times 10^{-3}$ eV$^2$, $\delta m^2\ll m^2$, and
$\sin^2\phi\lesssim \mathrm{few}\%$ \cite{Revi}. We assume the
supernova power-law density profile reported in Fig.~1 of
\cite{Sour}, and a representative neutrino energy $E=15$~MeV. The
electron neutrino survival probability, in the case of direct
hierarchy, is then given by
\begin{equation}\label{Peefig}
P_{ee}(\nu) \simeq [(1-P_E)\,
P_L(k_L,\omega+\epsilon_L)+P_E\,(1-P_L(k_L,\omega+\epsilon_L))]\,
e^{-2\pi r_H k_H \sin^2(\phi+\epsilon_H)}\ ,
\end{equation}
where we have neglected a small (additive) $O(\phi^2)$ term, the
main effect of $\phi$ being embedded in the exponential
suppression factor. In the above equation, $P_L$ is given by
Eq.~(\ref{PLeps}), while $P_E$ is calculated as in
Eqs.~(\ref{PEstd}) or (\ref{PEnonstd}). In the case of no Earth
crossing, $P_E$ is simply given by
\begin{equation}\label{PEnoEarth}
P_E=\sin^2\omega\ \ (\mathrm{no\ Earth\ effects})\ .
\end{equation}

Figure~1 shows curves of iso-$P_{ee}(\nu)$, calculated according
to Eq.~(\ref{Peefig}), with and without Earth
matter effects (dotted and solid lines, respectively),%
\footnote{In Fig.~1, Earth matter effects refer to a
representative case of mantle crossing, with pathlength $L=8500$
km, mass density $\rho=4.5$ g/cm$^3$, and electron number fraction
$Y_e=1/2$.}
in the plane of the $L$-transition variables $(\delta
m^2,\tan^2\omega)$. The three left panels refer to the case
$\tan^2(\phi+\epsilon_H)=0$ (corresponding to either
$\phi=\epsilon_H=0$ or to $\epsilon_H=-\phi$), which implies
$P_H=1$. The three right panels correspond instead to a nonzero
value for $\phi+\epsilon_H$, namely,
$\tan^2(\phi+\epsilon_H)=2\times 10^{-5}$, implying $P_H= 0.46$.
The comparison of left and right panels clearly shows the
suppression of $P_{ee}(\nu)$ due to $\phi+\epsilon_H\neq 0$.

In Fig.~1, from top to bottom, the three (left and right) panels
correspond to $\epsilon_L=0$, $\epsilon_L=+3\times 10^{-2}$, and
$\epsilon_L=-3 \times 10^{-2}$. The effect of nonzero $\epsilon_L$
is dramatic at small mixing angles, and leads to a distinctive and
well-known pattern for the isolines of $P_{ee}$
\cite{Ba91,Fo94,Be98,BeKa,Ku98}, which appear to be elongated
towards very small values of $\omega$ for positive $\epsilon_L$,
and split into two branches for negative $\epsilon_L$. In our
formalism, this behavior is simply captured in terms of the
$\cos^2(\omega+\epsilon_L)$ argument in $P_L$: for $\epsilon_L>0$
the argument increases monotonically for decreasing (small)
$\omega$, otherwise it first reaches a maximum at
$\omega=-\epsilon_L$ and then decreases. The Earth effect is also
similarly ``stretched'' in Fig.~1.

Let us consider in more detail the case of relatively large
$\omega$ [$\tan^2\omega\sim O(10^{\pm 1})$], which appears to be
generally favored by recent solar neutrino data \cite{Sola}. In
this case, independently of the value of $\delta m^2$, effects
induced by $\epsilon_L\neq 0$  are hardly recognizable in Fig.~1,
since the difference between $\omega$ and $\omega+\epsilon_L$ is
fractionally small, while the leading effect of nonstandard
interactions is embedded in the exponential suppression factor
($P_H$) of Eq.~(\ref{Peefig}).  In particular, for relatively
large values of $\delta m^2/E$ [providing the best fit to solar
neutrino data \cite{Sola} through the so called large mixing angle
(LMA) solution], it is $P_L\simeq 0$, and the $\nu_e$ survival
probability simply reads
\begin{equation}
P_{ee}\simeq P_E\; e^{-2\pi r_H k_H \sin^2(\phi+\epsilon_H)}\ \
(\mathrm{LMA\ solution})\ .
\end{equation}

Figure~2 shows the function $P_E(\delta m^2)$ for three
representative values of $\tan^2\omega$ in the LMA region. Other
parameters ($\epsilon_L$, neutrino energy and pathlength, mantle
density and electron number fraction) are as in Fig.~1. The case
of no Earth crossing [Eq.~(\ref{PEnoEarth})] is also shown as a
horizontal line in each panel. In general, Earth matter effects
appear to be relevant in all cases, producing a distinctive
oscillatory pattern in $P_E$. Different values of $\epsilon_L$
induce slightly different oscillation amplitudes in $P_E$, and
could thus (in principle) be distinguished, especially for large
values of $\delta m^2$, where the relative differences among the
curves in each panel increase.
 However, such discrimination would require very high
statistics observations. We conclude that, for mass-mixing
parameters in the LMA solution, and for typical supernova density
profiles and energy spectra, the main nonstandard effects are
associated with $\epsilon_H$ (through $P_H$), while smaller
effects are associated with $\epsilon_L$ (through $P_E$).

Concerning the main nonstandard effects,  the survival probability
in Eq.~(\ref{Peefig}) shows that, in general, supernova neutrinos
can be sensitive to values of $\tan^2(\phi+\epsilon_H)$ as low as
$\sim 10^{-5}$, corresponding to a (sub)percent sensitivity in
$\phi+\epsilon_H$. Indeed, in the standard case ($\epsilon_H=0$),
it has been shown that inversion of future galactic supernova data
might provide upper and lower bounds on $\phi$ in the percent
range \cite{Ba01}. Our results show that, in the presence of
additional nonstandard interactions $(\epsilon_H\neq 0)$, any
constraint about $\phi$ must actually be interpreted as a
constraint on $\phi+\epsilon_H$, namely, there is a strict
degeneracy between the vacuum mixing angle $\phi=\theta_{13}$ and
the nonstandard coupling $\epsilon_H$ given in Eq.~(\ref{epsH}).

The $(\phi,\epsilon_H)$ degeneracy in supernova neutrino
oscillations can be rather dangerous for the interpretation of
experimental data. Supernova $\nu$ oscillations may offer one of
the few opportunities to probe experimentally very small values of
$\phi$, and any uncontrolled bias of the kind
$\phi\to\phi+\epsilon_H$ may lead to dramatic differences in the
theoretical inferences from the experimental data. In extreme
cases, nonstandard interactions might either completely mimic
($\phi=0$, $\epsilon_H\neq 0$) or completely cancel
($\phi=-\epsilon_H$)
``standard'' $\phi$-related effects in the $H$ transition.%
\footnote{We remind the reader that, even assuming restrictive
bounds on the $\epsilon_{\alpha\beta}$
\protect\cite{BeKa,Be00,Be99,Gr00}, one cannot currently exclude
values of $\epsilon_H$ as large as a few percent.}
An analogous ``confusion scenario,'' with  degeneracy between
effects of nonstandard interactions and of $\phi\neq 0$, has
recently been discussed in \cite{Hu02} in the context of
oscillation searches at neutrino factories. The main message
emerging from such results is that constraining $\phi$ will
generally be a very challenging task, if allowance is made for
nonstandard (flavor changing) neutrino interactions, with strength
as weak as a few percent of the standard (flavor diagonal)
electroweak one.%
\footnote{At a more sophisticated level, for complex
$\epsilon_{\alpha\beta}$ there can also be confusion between
standard and nonstandard CP-violation phase effects in (very) long
baseline experiments; see, e.g.,  \protect\cite{Go01,Sa02}.}

We conclude this section with a positive remark, by showing that
not everything is necessarily biased by nonstandard interactions.
In particular, the spectrum hierarchy discrimination (direct {\em
vs\/} inverse) may still be viable. Let us consider the
phenomenologically interesting case of the so-called large mixing
angle solution (LMA) to the solar neutrino problem \cite{Sola},
requiring $\delta m^2\sim O(10^{-5})$ eV$^2$ and
$\tan^2\omega\lesssim 1$, so that $P_L\simeq 0$ (adiabatic $L$
transition). For small $\phi$, the supernova electron
(anti)neutrino survival probabilities are approximately given (up
to Earth matter effects) by
\begin{eqnarray}
P_{ee}^\mathrm{dir}(\nu)      &\simeq& \sin^2\omega\,
         e^{-2\pi r_H k_H  \sin^2(\phi+\epsilon_H)}\ ,\\
P_{ee}^\mathrm{dir}(\bar\nu)  &\simeq& \cos^2\omega\ ,\\
P_{ee}^\mathrm{inv}(\nu)      &\simeq& \sin^2\omega\ ,\\
P_{ee}^\mathrm{inv}(\bar\nu)  &\simeq & \cos^2\omega\,
    e^{-2\pi r_H k_H \sin^2(\phi+\epsilon_H)}\ ,
\end{eqnarray}
where the superscripts denote the cases of direct or inverse
hierarchy. From such equations one can derive the following
$\epsilon_H$-independent inequalities,
\begin{eqnarray}
  P_{ee}^\mathrm{dir}(\nu) &<& P_{ee}^\mathrm{inv}(\nu) \ , \\
  P_{ee}^\mathrm{dir}(\bar\nu)&>&P_{ee}^\mathrm{inv}(\bar\nu)\ ,
\end{eqnarray}
and
\begin{equation}
  P_{ee}^\mathrm{dir}(\nu)/P_{ee}^\mathrm{dir}(\bar\nu) <
  P_{ee}^\mathrm{inv}(\nu)/P_{ee}^\mathrm{inv}(\bar\nu) \ ,
\end{equation}
which become all stronger for decreasing neutrino energy. Such
relations can then be used to infer information about the mass
spectrum hierarchy from general features of supernova
(anti)neutrino event
spectra,%
\footnote{The phenomenological implementation of the above
inequalities requires, of course, that $P_{ee}$ is folded with
$\nu$ (or $\overline\nu$) energy spectra and cross sections. This
further step is beyond the scope of the present work.}
analogously to the standard case \cite{Mi01,Lu01,Wood}. The
hierarchy discrimination becomes impossible, however, for
$\sin^2(\phi+\epsilon_H) \ll 10^{-5}$ (corresponding to
$\exp[-2\pi r_H k_H \sin^2(\phi+\epsilon_H)]\simeq1$), in which
case the above three relations become equalities.

We finally remind that, under certain conditions (see the end of
Sec.~II), additional nonstandard transitions might take place just
above the neutrinosphere \cite{Va87,Vall}. In this case (not
considered here), the global $P_{ee}$ function should be obtained
by convolving such transitions with those ($H$ and $L$) considered
in this and in the previous section, leading to a more complicated
phenomenology.

\section{Summary}

We have revisited the effects of nonstandard four-fermion
interactions (with strength $\epsilon_{\alpha\beta} G_F$) on
supernova neutrino oscillations occurring above the protoneutron
star. Under reasonable approximations, we have found that the
oscillation probability can be written in an analytical form,
valid in the whole $2\nu$ parameter space, and applicable also to
the $3\nu$ parameter space in hierarchical (direct or inverse)
cases. Our approach generalizes previous results, which are
recovered as specific limits. We find that, as far as the
transitions at high $(H)$ and low $(L)$ density are concerned, the
main effects of the new interactions can be embedded through
(positive or negative) shifts of the relevant mixing angles
$\phi=\theta_{13}$
 and $\omega=\theta_{12}$, namely,
$\phi\to\phi+\epsilon_H$ and $\omega\to\omega+\epsilon_L$,
 respectively [see Eqs.~(\ref{epsH}) and (\ref{epsL})]. Barring the case of
small $\omega$ (disfavored by solar neutrino data), and apart from
small $\epsilon_L$-induced Earth matter effects, the main
phenomenological implication of such results is a strict
degeneracy between standard $(\phi)$ and nonstandard
$(\epsilon_H)$ effects on the $H$ supernova $\nu$ transition.
However, such a degeneracy does not necessarily spoil the
discrimination between direct and inverse neutrino mass spectrum
hierarchy. Our work is complementary to studies \cite{Vall} of
nonstandard transition phenomena possibly occurring just above the
neutrinosphere.

\begin{acknowledgments}
This work was supported in part by the Italian  {\em Istituto
Nazionale di Fisica Nucleare\/} (INFN) and {\em Ministero
dell'Istruzione, dell'Universit\`a e  della Ricerca\/} (MIUR)
under the  project ``Fisica Astroparticellare.'' We thank J.W.F.\
Valle for very useful discussions.
\end{acknowledgments}



\begin{figure}
\includegraphics[width=13cm]{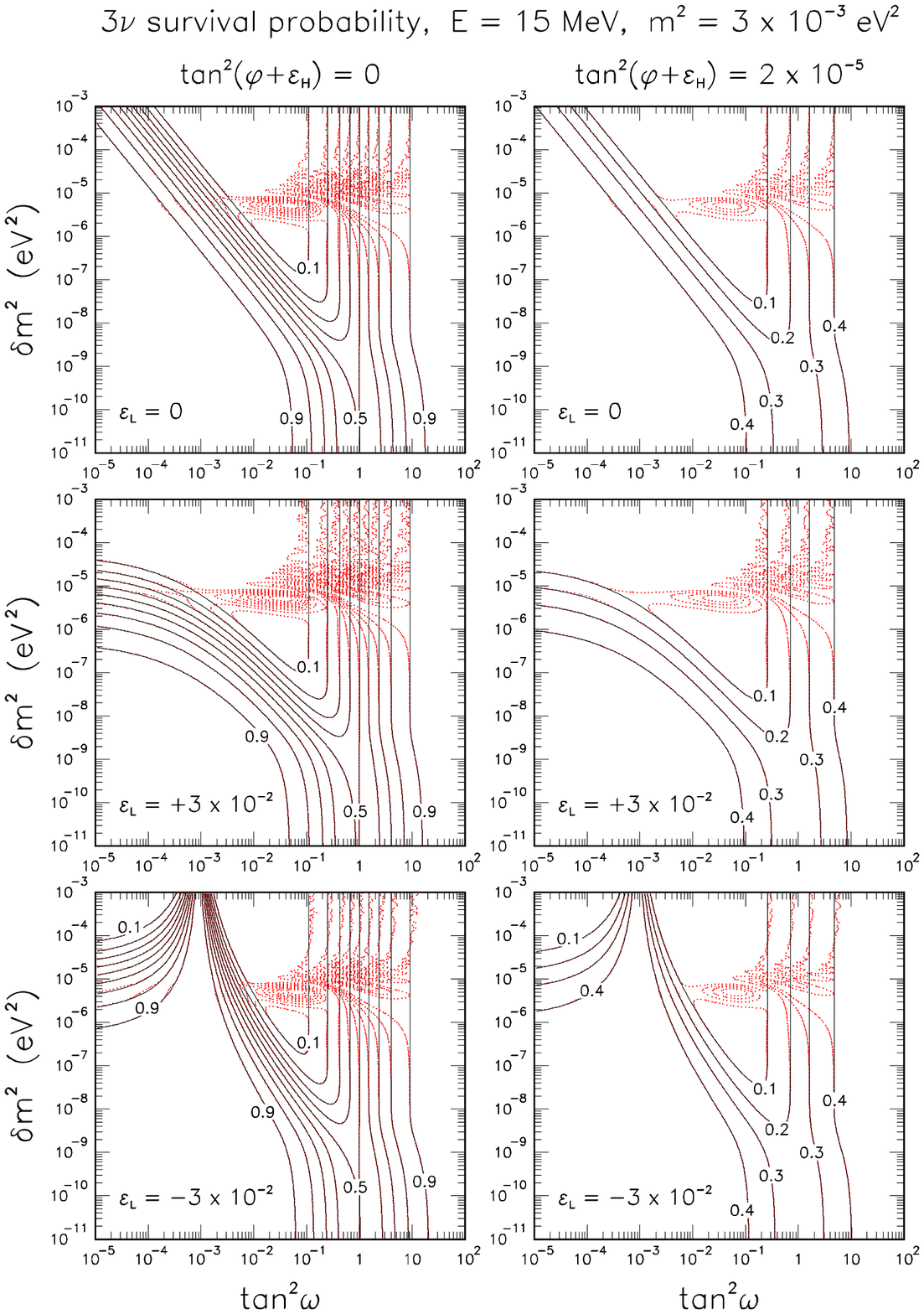}
\caption{Isolines of $P_{ee}$ for supernova neutrinos with direct
mass hierarchy, in the plane of the $L$-transition variables
$(\delta m^2,\tan^2\omega)$, with (dotted lines) and without
(solid lines) Earth effects (calculated for $L=8500$ km,
$\rho=4.5$ g/cm$^3$, and $Y_e=1/2$). Nonstandard interactions are
parameterized through the dimensionless couplings
$\epsilon_{L,H}$. The values of $m^2$ and $E$ are fixed at
$3\times 10^{-3}$ eV$^2$ and 15 MeV, respectively. Left (right)
panels refer to $\tan^2(\phi+\epsilon_H)$=0 ($=2\times 10^{-5}$).
The parameter $\epsilon_L$ takes the representative values $0$
(upper panels), $+3\times 10^{-2}$ (middle panels), and $-3\times
10^{-2}$ (lower panels). See the text for details.}
\end{figure}
\begin{figure}
\includegraphics[width=12cm]{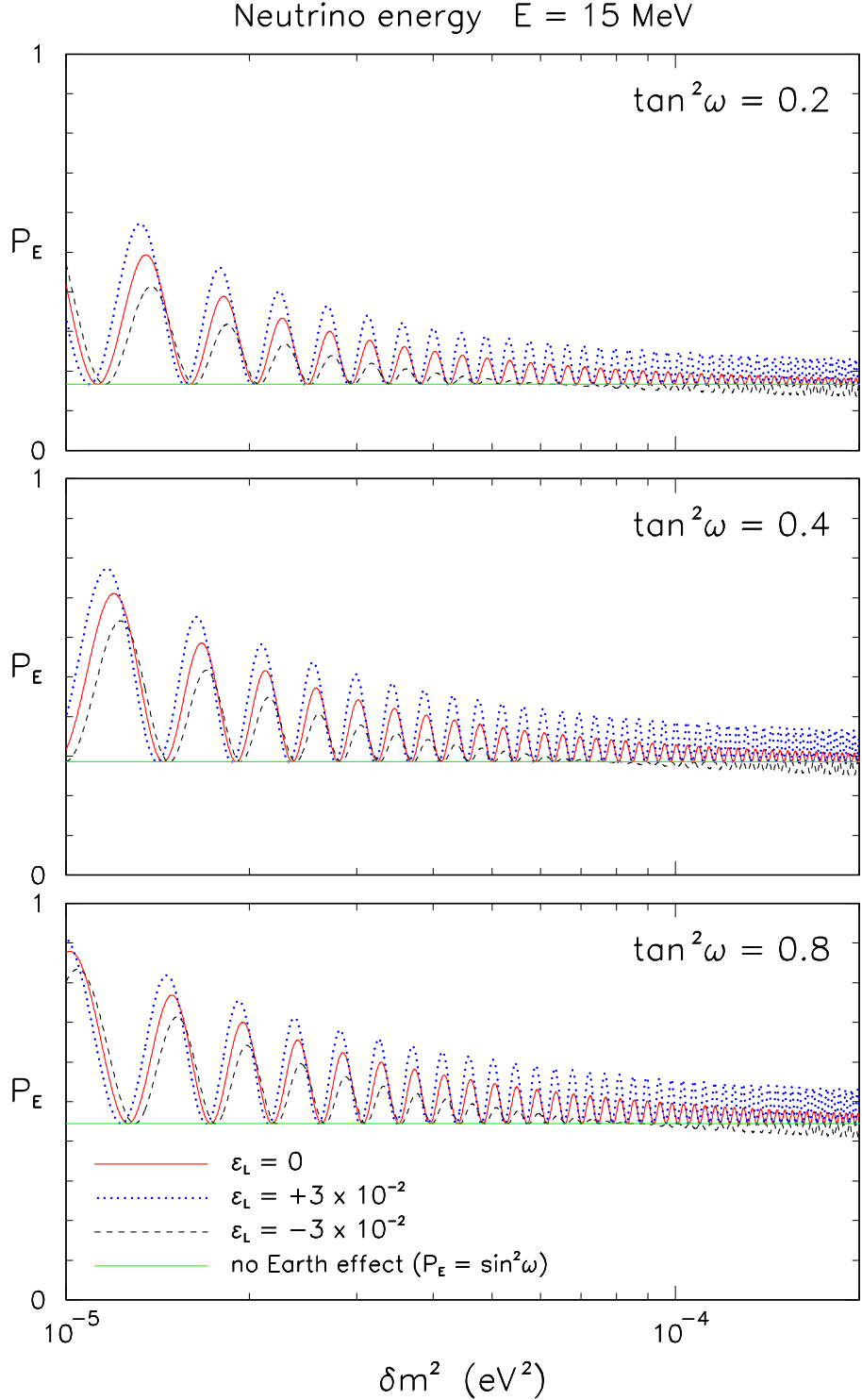}
\caption{Earth matter effects: Curves of the probability
$P_E=P(\nu_2\to\nu_e)$ as a function of $\delta m^2$ for mantle
crossing of supernova neutrinos (with $E$, $L$, $\rho$ and $Y_e$
as in Fig.~1). The chosen range of $\delta m^2$ and the three
representative values of $\tan^2\omega$ refer to the so-called LMA
solution to the solar neutrino problem \protect\cite{Sola}.
Nonstandard interactions affect the amplitude and the phase of the
oscillatory pattern through $\epsilon_L$. The horizontal solid
line represents the case of no Earth effect.}
\end{figure}
\end{document}